\RequirePackage{ifpdf}
\ifpdf 
\documentclass[pdftex]{sigma}
\else
\documentclass{sigma}
\fi

\usepackage{xypic}

\begin{document}

\allowdisplaybreaks

\renewcommand{\thefootnote}{$\star$}

\renewcommand{\PaperNumber}{054}

\FirstPageHeading

\ShortArticleName{Eigenvalues for the Nondegenerate Three-Dimensional Kepler--Coulomb Potential}

\ArticleName{Algebraic Calculation of the Energy Eigenvalues\\ for the Nondegenerate Three-Dimensional\\ Kepler--Coulomb Potential\footnote{This paper is a
contribution to the Special Issue ``Symmetry, Separation, Super-integrability and Special Functions~(S$^4$)''. The
full collection is available at
\href{http://www.emis.de/journals/SIGMA/S4.html}{http://www.emis.de/journals/SIGMA/S4.html}}}

\Author{Yannis TANOUDIS and Costas DASKALOYANNIS}

\AuthorNameForHeading{Y.~Tanoudis and C.~Daskaloyannis}

\Address{Mathematics Department, Aristotle University of Thessaloniki, 54124 Greece}

\Email{\href{mailto:tanoudis@math.auth.gr}{tanoudis@math.auth.gr}, \href{mailto:daskalo@math.auth.gr}{daskalo@math.auth.gr}}

\ArticleDates{Received February 01, 2011, in f\/inal form May 22, 2011;  Published online June 03, 2011}

\Abstract{In the three-dimensional f\/lat space, a classical Ha\-mi\-lton\-ian, which  has f\/ive functionally independent integrals of motion, including the Hamiltonian, is characterized as superintegrable. Kalnins, Kress and Miller ({\it  J.~Math. Phys.}  {\bf  48} (2007), 113518,  26~pages) have proved that, in the case of nondegenerate potentials, i.e.\ potentials depending linearly on four parameters, with quadratic symmetries, posses a sixth quadratic integral, which is linearly independent of the other integrals. The exi\-stence of this sixth integral imply that the  integrals of motion form  a ternary quadratic Poisson algebra with f\/ive generators. The superintegrability of the generalized Kepler--Coulomb potential that was investigated by Verrier and Evans ({\it J.~Math. Phys.}  {\bf 49} (2008),  022902, 8~pages) is a special case of superintegrable system, having two independent integrals of motion of fourth order among the remaining quadratic ones.
The corresponding Poisson algebra of integrals is  a quadratic one, having the same special form, characteristic to the nondegenerate case of systems with quadratic integrals. In this paper, the ternary quadratic associative algebra corresponding to the quantum Verrier--Evans system is discussed. The subalgebras structure, the Casimir operators and the  the f\/inite-dimensional representation of this algebra are studied and the energy eigenvalues of the nondegenerate Kepler--Coulomb are calculated.}


\Classification{81R12; 37J35; 70H06; 17C90}

\section{Introduction}\label{int1}

 In the $N$-dimensi\-onal space one Hamiltonian system characterized as superintegrable if it
has $2 N-1$ integrals.

Kalnins, Kress and Miller have studied \cite{KalKrMi07,KalKrMi07JPA} three-dimensional  superintegrable systems, whose the potentials depend on four constants; these systems are  referred  as  \emph{nondegenerate potentials}. The case in which one three-dimensional potential has fewer ``free parameters'' than four, def\/ines a  potential, which is called  \emph{degenerate potential}. Verrier and Evans \cite{VerEvans08} have introduced a new classical superintegrable Hamiltonian,
\begin{gather*}
H=\frac{1}{2} \left(p_{x}^2+p_{y}^2+p_{z}^2\right)-\frac{k }{ \sqrt{x^2+y^2+z^2}}+\frac{k_1}{x^2}+\frac{k_2}{y^2}+\frac{k_3}{z^2},
\end{gather*}
which is a nondegenerate  generalized Kepler--Coulomb system.
The above Hamiltonian is a~superintegrable system with quadratic and quartic, in momenta, integrals of motion. The quartic integrals are generalizations of the Laplace--Runge--Lenz vectors of the ordinary Kepler--Coulomb potential.

The quantum form of the nondegenerate Kepler--Coulomb  Hamiltonian is
\begin{gather}
H=-\frac{\hbar^2}{2}(\partial_{xx}+\partial_{yy}+\partial_{zz})-\frac{\hbar^2  \mu }{  \sqrt{x^2+y^2+z^2}}\nonumber\\
\hphantom{H=}{} +\frac{ \hbar^2 (4 \mu_1^2-1)}{8 x^2}+\frac{ \hbar^2 (4 \mu_2^2-1)}{8 y^2} +\frac{\hbar^2 (4  \mu_3^2-1)}{8 z^2}.
\label{eq:Quantum non degenerate mu}
\end{gather}

Kalnins, Williams, Miller and Pogosyan in \cite{KaWiMiPo99} have studied  the energy eigenvalues using the method of separation of variables for  potentials, one among them is the generalized Kepler--Coulomb system.
 In this paper the energy eigenvalues of the nondegenerate generalized Kepler--Coulomb system are calculated  by using  algebraic methods. In Section~\ref{sec:TwoDim} we recall the method of calculation of energy eigenvalues using quadratic ternary algebras~\cite{Das01}. In Section~\ref{CoulombQuadraticAlgebra} the structure of the algebra generated by the integrals of the nondegenerate Kepler--Coulomb system is studied.  In Section~\ref{sec:Energy} we apply the method of Section~\ref{sec:TwoDim} and we calculate the energy eigenvalues of the nondegenerate Coulomb system.
The commutation relations of the quadratic algebra are given in~\cite{Dubna09}, for clarity reasons these relations are reproduced in the Appendix.

\section{Quadratic algebra for two-dimensional quantum\\ superintegrable systems with quadratic integrals}\label{sec:TwoDim}

Any two-dimensional quantum superintegrable system with integrals quadratic in momenta is described by the Hamiltonian $H$ and two functionally independent integrals of motion~$A$ and~$B$. The integrals~$A$ and~$B$ commute with Hamiltonian $H$, but they don't commute between them
\[
\left[H,  A\right]= 0, \qquad \left[H,  B\right]= 0  \qquad \mbox{and}\qquad \left[A, B\right]\ne 0.
\]
The above relations can be presented by the following diagram
\begin{gather}\label{eq:diagram}
\xymatrix{ A \ar@{--}[r] & H \ar@{--}[r] & B},
 \end{gather}
 where the dashed line joining two operators  means that the corresponding commutator is zero, the absence of any joining line between $A$ and $B$ means that the corresponding commutator is dif\/ferent to zero

Let $\mathcal{A}=\mathbb{C}\langle A,B,H\rangle$ be the unital algebra generated by the operators $A$, $B$, $H$, the generators of this algebra satisfy  ternary relations, which are quadratic extensions of the  enveloping algebra of a Lie triple system \cite{Jac51}
\begin{gather}
\left[ A,\left[A,B\right] \right ]=\alpha A^2+\beta B^2+ \gamma \{ A,B \}+\delta A+ \epsilon B +\zeta ,\nonumber\\
 \left [B,  \left[A,B\right] \right ] = a A^2-\gamma B^2-\alpha  \{ A, B\}  +d
A-\delta B+z.\label{eq:QLTS}
 \end{gather}
In a Lie triple system~\cite{Jac51} in the right hand side of the above equations  are only linear functions of the operators~$A$ and~$B$. In the case of superintegrable systems with quadratic integrals of motion there are also quadratic terms.  Some of the coef\/f\/icients of the ternary quadratic algebra~(\ref{eq:QLTS}) depend generally on the energy~$H$
\begin{gather}
 \delta=\delta_{1} H+\delta_{0}, \qquad
 \epsilon=\epsilon_{1} H+\epsilon_{0}, \nonumber\\
 \zeta=\zeta_{2} H^2+\zeta_{1} H+\zeta_{0}, \qquad
 d=d_{1} H+d_{0}, \qquad
 z=z_{2} H^2+z_{1} H+z_{0}.\label{eq:QLTScoef}
\end{gather}
 We are interested to calculate the energy eigenvalues of the operator $H$,  therefore we search to calculate the values of the energy corresponding to f\/inite-dimensional representations of the algebra~(\ref{eq:QLTS}).

There is a  Casimir $[K,A]=[K,B]=0$  and  $K=K(H)$:
\begin{gather}
  K=\left[A,B\right]^2-\alpha \{ A^2,B\}-\gamma\{ A,B^2 \}+\left(\alpha
\gamma -\delta+\frac{a \beta}{3}\right)\{A,B\}-\frac{2 \beta}{3} B^3
\nonumber\\
\hphantom{K=}{}
+\left(\gamma^2-\epsilon-\dfrac{\alpha \beta}{3}\right) B^2+ \left(-\gamma
\delta+2\zeta-\frac{\beta  d}{3}\right)B   +\frac{2 a}{3} A^3 \nonumber\\
\hphantom{K=}{}
+\left(d+\frac{a
\gamma}{3}+\alpha^2\right)
A^2+\left(\frac{a \epsilon}{3}+\alpha \delta+2 z\right) A
=
 h_0  + h_1 H + h_2 H^2 + h_3 H^3.\label{eq:QLTSCas}
\end{gather}
In~\cite{DasYps06} the classical two-dimensional superintegrals with quadratic integrals of motion satisfy a~Poisson quadratic algebra.
All the quantum superintegrable systems with quadratic integrals of motion satisfy the algebra (\ref{eq:diagram})--(\ref{eq:QLTSCas})~\cite{DasTan07}. The classical three-dimensional superintegrable systems with quadratic integrals on a f\/lat space have a~structure of ternary quadratic algebra~\cite{TaDas09}.

In~\cite{Das01}  the unitary representation of this algebra is studied.  In the case of $\gamma\ne 0$ and $\beta=0$, the eigenvalues of the operator $A$ are given by the formula
\begin{gather*}
A(x)=\frac{\gamma}{2} \biggl((x+u)^2 -\frac 14-\frac{\epsilon}{\gamma^2}\biggr),
\end{gather*}
where $x=0,1,2,\ldots$. The energy eigenvalues of the operator $H$ with degeneracy equal to $p+1$, where $p=0,1,2,\ldots$ are determined by solving the system of equations
\begin{gather*}
\Phi(0,u)=0,\qquad \Phi(p+1,u)=0,\qquad  \mbox{and}\qquad  \Phi(x,u)>0, \qquad \mbox{where}\quad
x=1,2,\ldots, p.
\end{gather*}
The structure function is given by the next relation
\begin{gather}
\Phi(x)=-3072 K (2 (u+x)-1)^2 \gamma ^6-48 (2 (u+x)-3) (2 (u+x)-1)^4
\nonumber\\
\phantom{\Phi(x)=}{}\times
(2 (u+x)+1) \gamma ^6 \left(\epsilon  \alpha ^2-\gamma  \delta  \alpha -d \gamma ^2+a \gamma  \epsilon \right) + (2 (u+x)-3)^2 (2 (u+x)-1)^4 \nonumber\\
\phantom{\Phi(x)=}{}
\times (2 (u+x)+1)^2 \left(3 \alpha ^2+4 a \gamma \right) \gamma ^8 + 768 \left(4 \zeta  \gamma ^2-2 \delta  \epsilon  \gamma
   +\alpha  \epsilon ^2\right)^2 + 32 (2 (u+x)-1)^2
\nonumber\\
\phantom{\Phi(x)=}{}\times
    \left(12 (u+x)^2-12 (u+x)-1\right) \gamma ^4 \left(8 z \gamma ^3+2 \delta ^2 \gamma ^2-4 d \epsilon  \gamma ^2+4
   \alpha  \zeta  \gamma ^2+2 a \epsilon ^2 \gamma \right.
\nonumber\\
\left.\phantom{\Phi(x)=}{}
   -6 \alpha  \delta  \epsilon  \gamma +3 \alpha ^2 \epsilon ^2\right)-256 (2 (u+x)-1)^2 \gamma ^2 \left(-4 z \gamma ^5+2 \delta ^2 \gamma ^4+2 d \epsilon  \gamma ^4+4 \alpha  \zeta  \gamma ^4\right.
\nonumber\\
\left.\phantom{\Phi(x)=}{}
   +12
   z \epsilon  \gamma ^3-12 \delta  \zeta  \gamma ^3-3 d \epsilon ^2 \gamma ^2+6 \delta ^2 \epsilon  \gamma ^2+12 \alpha  \epsilon
   \zeta  \gamma ^2+a \epsilon ^3 \gamma -9 \alpha  \delta  \epsilon ^2 \gamma +3 \alpha ^2 \epsilon ^3\right).\label{eq:PhiDef}
\end{gather}
In~\cite{Das01} this method is used for the calculation of energy eigenvalues for the two-dimensional superintegrable systems on the plane. This quadratic, cubic and generally polynomial algebras is the subject of current investigations.  The cubic extension of the above algebra for superintegrable systems with an integral of motion cubic in momenta is studied in a series of papers by I.~Marquette and P.~Winternitz~\cite{MarWin07JMP48,MarWin08JPA41}, by I.~Marquette \cite{Mar09JMP50a,Mar09JMP50b,Mar09JMP50c}. The case of higher order integrals of motion can be found in~\cite{Mar10JPA43}, the case of one-dimensional position-dependent mass Schr\"odinger equation in~\cite{Quesne07_SIGMA}. These methods were applied to three-dimensional MICZ-Kepler system in \cite{Mar10JMP51}.

\section[Quadratic algebra for the nondegenerate Kepler-Coulomb system]{Quadratic algebra for the nondegenerate\\ Kepler--Coulomb system}\label{CoulombQuadraticAlgebra}

The  nondegenerate Kepler--Coulomb system (\ref{eq:Quantum non degenerate mu}) possesses three quadratic integrals which denoted by $A_1$, $A_2$, $B_2$ and one integral of fourth order in addition to last mentioned quadratic ones, which denoted by~$B_1$
\begin{gather*}
A_1=\frac{1}{2} J^2+\frac{\hbar^2 (4 \mu_1^2-1)(y^2+z^2)}{8 x^2}+\frac{\hbar^2 (4 \mu_2^2-1) (x^2+z^2)}{8 y^2}+\frac{\hbar^2 (4 \mu_3^2-1)(x^2+y^2)}{8 z^2},
\\
A_2=\frac{1}{2} J_{3}^2+\frac{\hbar^2 (4 \mu_1^2-1) y^2}{8 x^2}+\frac{\hbar^2 (4 \mu_2^2-1) x^2}{8 y^2},
\\
B_2=\frac{1}{2} J_2^2+\frac{\hbar^2 (4 \mu_1^2-1) z^2}{8 x^2}+\frac{\hbar^2 (4 \mu_3^2-1) x^2}{8 z^2},
\\
p_x=- i \hbar \partial_x, \qquad p_y=- i \hbar \partial_y ,\qquad p_z=- i \hbar \partial_z,\qquad J_1=i \hbar (z \partial_y-y\partial_z),
\\
  J_2=i \hbar (x \partial_z-z\partial_x), \qquad J_3= i \hbar (y \partial_x-x\partial_y), \qquad J=J_1^2+J_2^2+J_3^2,
\\
B_1=
 \left ( \frac{1}{2} \{ J_1,p_y\}-\frac{1}{2} \{J_2,p_x\}-2 z \left(\frac{-\hbar^2 \mu}{2\sqrt{x^2+y^2+z^2}}+\frac{\hbar^2 (4 \mu_1^2-1)}{8 x^2}+\frac{\hbar^2 (4 \mu_2^2-1)}{8 y^2}\right.\right.\\
 \left.\left.
{}
 +
\frac{\hbar^2 (4 \mu_3^2-1)}{8 z^2} \right) \right)^2 +
\left\{\frac{1}{4} \left({\{x,p_x\}+\{y,p_y\}+\{z,p_z\}}\right)^2,\frac{ \hbar^2 (4 \mu_3^2-1)}{8 z^2} \right\}
+\frac{5 \hbar^4 (4 \mu_1^2-1) }{16 z^2}.
\end{gather*}
The above operators satisfy the following zero commutation relations
\begin{gather}\label{eq:zero1}
[H, A_i]=0, \qquad [H, B_i]=0, \qquad [A_1,B_2]=0, \qquad [A_2,B_1]=0.
\end{gather}
According to  Kalnins--Kress--Miller ``5 to 6''  theorem \cite{KalKrMi07}, in any three-dimensional Hamiltonian system with nondegenerate potential with quadratic integrals of motion, there is always exist a~$6$th integral that is functionally depended with the other integrals. This sixth integral of motion is linearly independent with the 5 functionally independent integrals.  In the case of Kepler--Coulomb potential the last mentioned sixth integral is an integral of fourth order in momenta similar to  the integral $B_1$,  given by the following formula
\begin{gather*}
F=  \left ( -\frac{1}{2} \{ J_1,p_z\}+\frac{1}{2} \{J_3,p_x\}\right. \\
\left.
\phantom{F=}{}-2 y \left( \frac{-\hbar^2  \mu}{2 \sqrt{x^2+y^2+z^2}}+\frac{\hbar^2 (4 \mu_1^2-1)}{8 x^2}+\frac{\hbar^2 (4 \mu_2^2-1)}{8 y^2}+\frac{\hbar^2 (4 \mu_3^2-1)}{8 z^2}\right )\right)^2\\
\phantom{F=}{}
 +
 \left\{\frac{1}{4}\left({ \{x,p_x\}+\{y,p_y\}+\{z,p_z\}}\right)^2,\frac{ \hbar^2 (4 \mu_2^2-1) }{8 y^2} \right\}+\frac{5 \hbar^4 (4 \mu_2^2-1) }{16 y^2}.
\end{gather*}
This integral satisf\/ies the following commutation relations
\begin{gather}\label{eq:zero2}
[H,F]=0,\qquad [F,B_2]=0.
\end{gather}
The graph representing the zero commutation relations (\ref{eq:zero1}) and (\ref{eq:zero2}) is the following one
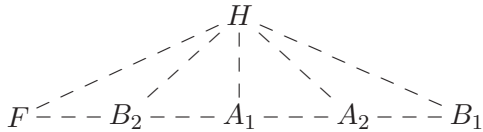
\begin{gather*}
\xymatrix{                  &   &H\ar@{--}[lld]\ar@{--}[ld]\ar@{--}[d]\ar@{--}[rd]\ar@{--}[rrd]
\\ F  \ar@{--}[r] & B_2  \ar@{--}[r] &A_1 \ar@{--}[r] & A_2 \ar@{--}[r] & B_1}
 \end{gather*}
In Appendix~\ref{appendix} all the non zero ternary relations are given. By inspecting the relations~(\ref{eq:alg1a}) and~(\ref{eq:alg1b}) in the Appendix, we can see that the unital algebra generated by the operators $A_1$, $B_1$, $A_2$, $H$ corresponds to the graph:
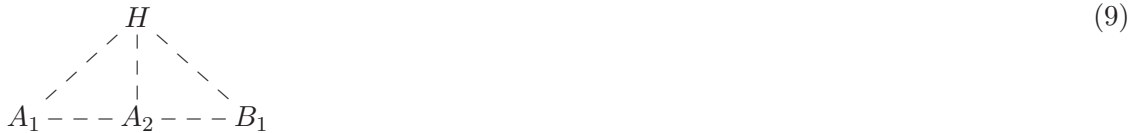
\begin{gather}\label{eq:graph1}
\xymatrix{                     &H\ar@{--}[ld]\ar@{--}[d]\ar@{--}[rd]
\\ A_1  \ar@{--}[r] &A_2 \ar@{--}[r] & B_1}
 \end{gather}
This is  a quadratic subalgebra  corresponding to some quadratic algebra of the form given by equations (\ref{eq:QLTS})--(\ref{eq:QLTSCas}) but the coef\/f\/icients (\ref{eq:QLTScoef}) depend on the operators $H$ and $A_2$
\begin{gather}
\alpha= -16\hbar^2 H, \qquad \gamma=4\hbar^2, \qquad \delta= 16 \hbar^2 A_2 H - 2\hbar^4 (4\mu_1^2+4\mu_2^2+12\mu_3^2) H -4 \hbar^6 \mu^2,
\nonumber\\
\epsilon= 2\hbar^4 (2\mu_1^2+2\mu_2^2+2\mu_3^2-3),\qquad d=16 \hbar^4 (5-4\mu_3^2) H^2, \qquad a=0,
\nonumber\\
z=-32\hbar^4 A_2 H^2 + 4 \hbar^8 \mu^2 (3-4\mu_3^2) H \nonumber\\
\phantom{z=}{} + 2 \hbar^6 (12\mu_1^2+12\mu_2^2-16\mu_3^4-8(2\mu_1^2+2\mu_2^2-3)-1) H^2,
\nonumber\\
\zeta=4\hbar^6 \mu^2 A_2 + 2\hbar^6 \mu^2 (1-2\mu_3^2)+2 (4\mu_1^2+4\mu_2^2+4\mu_3^2) A_2 H \nonumber\\
\phantom{\zeta=}{} +\hbar^6 (1-2\mu_3^2(4\mu_1^2+4\mu_2^2+4\mu_3^2-5)) H.\label{eq:coef1}
\end{gather}
Moreover, the above quadratic subalgebra possesses a Casimir invariant given from the following expression
\begin{gather*}
K_1=4 \mu ^4 \left(\mu _3^2-1\right) \hbar ^{12}+4 \hbar ^{10}  \mu ^2 \left(4 \mu _3^4-23 \mu _3^2+4 \mu _1^2 \left(\mu _3^2-1\right)+4 \mu _2^2 \left(\mu _3^2-1\right)+8\right) H\\
\hphantom{K_1=}{}
   +56 \hbar ^8 \mu ^2 A_2 H  +\hbar ^8 \left(16 \left(\mu _3^2-1\right) \mu _1^4+4 \left(8 \mu _3^4-34 \mu _3^2+8 \mu _2^2 \left(\mu
   _3^2-1\right)+5\right) \mu _1^2+97 \mu _3^2\right.\\
\left.\hphantom{K_1=}{}
   +4 \left(4 \mu _3^6-42 \mu _3^4+4 \mu _2^4 \left(\mu _3^2-1\right)+\mu _2^2 \left(8 \mu _3^4-34 \mu
   _3^2+5\right)\right)+15\right) H^2  \\
\hphantom{K_1=}{}
   +4\hbar ^6 \left(28 \mu _1^2+28 \mu _2^2+52 \mu _3^2-31\right) A_2 H^2 -48 \hbar ^4 A_2^2 H^2.
\end{gather*}
We must remark that  Marquette \cite{Mar10JMP51} studying the  MICZ-Kepler system has found quadratic algebras with coef\/f\/icients, which depend of two commuting operators.

By inspecting the relations (\ref{eq:alg2a}) and (\ref{eq:alg2b}) in the Appendix, we can see that the unital algebra generated by the operators $A_2$, $B_2$, $A_1$ corresponds to the graph
\begin{gather}\label{eq:graph2}
\xymatrix{                     
B_2  \ar@{--}[r] &A_1 \ar@{--}[r] & A_2}.
 \end{gather}
This is  a quadratic subalgebra  corresponding to some quadratic algebra of the form given by equations (\ref{eq:QLTS})--(\ref{eq:QLTSCas})  the coef\/f\/icients (\ref{eq:QLTScoef}) depend on the operator $A_1$
\begin{gather}
\alpha=\gamma=4\hbar^2, \qquad \delta=-4\hbar^2 A_1 + \hbar^4(2\mu_1^2-3), \qquad \epsilon=2\hbar^4 (2\mu_1^2+2\mu_2^2-3),
\nonumber\\
\zeta= \hbar^6 (\mu_2^2-\mu_1^2)+ \hbar^4 (3-4 \mu_1^2) A_1, \qquad \beta=a=0, \qquad d=-2\hbar^4 (2\mu_1^2+2\mu_2^2-3),
\nonumber\\
z=\hbar^6 (\mu_1^2-\mu_3^2)+ \hbar^4 (4\mu_1^2-3) A_1.\label{eq:coef2}
\end{gather}
This subalgebra possesses a Casimir invariant that it is given as follows
\begin{gather*}
K_2=\frac{1}{8} \left(\left(-32 \left(\mu _3^2-1\right) \mu _2^2+32 \mu _3^2+2\right) \mu _1^2-30 \mu _3^2+\mu _2^2 \left(32 \mu _3^2-30\right)+9\right) \hbar
   ^8\\
   \hphantom{K_2=}{}
+\frac{3}{2} \hbar ^6  \left(12 \mu _1^2-7\right) A_1 +4 \hbar ^4\left(\mu _1^2-1\right) A_1^2.
\end{gather*}
The relations (\ref{eq:alg3a}) and (\ref{eq:alg3b}) in the Appendix imply that the quadratic subalgebra corresponding to the following diagram
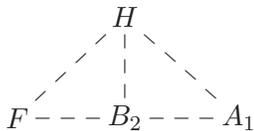
\begin{gather*}
\xymatrix{                     &H\ar@{--}[ld]\ar@{--}[d]\ar@{--}[rd]
\\ F  \ar@{--}[r] &B_2 \ar@{--}[r] & A_1}
 \end{gather*}
is a quadratic subalgebra of the form  (\ref{eq:QLTS})--(\ref{eq:QLTSCas}).

\section{Calculation of the energy eigenvalues}\label{sec:Energy}

Using the theory given in Section~\ref{sec:TwoDim} for the subalgebra generated by the operators $A_2$, $B_2$, $A_1$  (see graph~(\ref{eq:graph2}) and the coef\/f\/icients given by  equation~(\ref{eq:coef2})), the structure function~$\Phi(u,x)$ in equation~(\ref{eq:PhiDef})  is written
\begin{gather*}
\Phi(u, x)=3 \cdot 2^{18} \hbar ^{16} \left(2 (u+ x)-\mu _1-\mu _2-1\right)
   \left(2 (u+ x)+\mu _1-\mu _2-1\right) \\
   \hphantom{\Phi(u, x)=}{}
   \times
   \left(2 (u+
   x)-\mu _1+\mu _2-1\right) \left(2 (u+ x)+\mu _1+\mu
   _2-1\right) \\
   \hphantom{\Phi(u, x)=}{}
   \times
   \left(8 \hbar ^2 (u+x)^2- 8\hbar^2 (\mu_3+1)(u+x)-2 \hbar^2 \biggl (\mu
   _1^2  + \mu _2^2 -\frac{1}{2} \biggr ) \right.\\
\left.\hphantom{\Phi(u, x)=}{}
   +4 \hbar ^2 \biggl( \mu _3 +\frac{1}{2} \biggr)-4
   A_1\right) \bigg(8 \hbar ^2 (u+x)^2+ 8 \hbar^2 (\mu_3-1)(u+x)\\
\hphantom{\Phi(u, x)=}{}
 -2 \hbar^2 \left(\mu _1^2 + \mu _2^2 -\frac{1}{2} \right)  -4 \hbar ^2 \biggl(\mu _3-\frac{1}{2} \biggr ) -4
   A_1\bigg).
\end{gather*}
The value of parameter $u$ corresponding to the representation of the ternary algebra of dimension $p+1$  as well the eigenvalues of the operator $A_1$ determined by the next relations
\[
\Phi(u,0)=\Phi(u,p+1)=0.
\]
Since the structure function is a positive for $x=1,2,\ldots,p$  the values of $u$ and the corresponding eigevalues of $A_2$ and  $ A_1$ can be calculated analytically.

\noindent \textbf{Class I.}
\begin{gather*}
u=\frac{1}{2}+\frac{\mu_1+\mu_2}{2},
\\
2 A_2 (x) = \hbar^2 M^2-\hbar^2 \left( \mu_1^2+\mu_2^2\right)+\frac{\hbar^2}{2},
\end{gather*}
where $M=2 x + \mu_1+ \mu_2 + 1 $,  $x=0, 1, \ldots, p$,
\begin{gather*}
\Phi(x)=3 \hbar^{20}  2^{28} x (p-x+1)(x+ \mu_1)(x+\mu_2) (x+\mu_1+ \mu_2)(p-x+\mu_3+1)   \\
\phantom{\Phi(x)=}{} \times (p+x+\mu_1+\mu_2+1)(p+x+\mu_1+\mu_2+\mu_3+1).
\end{gather*}

\noindent \textbf{Class II.}
\begin{gather*}
 u=-\frac{1}{2}(1+2 p)-\frac{\mu_1+\mu_2}{2},
\\
2 A_2(x)
= \hbar^2 M^2 - \hbar^2 \left( \mu_1^2+\mu_2^2\right)+\frac{\hbar^2}{2},
\end{gather*}
where $M= 2 (p-x) + \mu_1+ \mu_2 + 1 $,
\begin{gather*}
\Phi(x)= 3 \hbar^{20}  2^{28} x (p-x+1)(p-x+\mu_1+1)(p-x+\mu_2+1)(2p-x+\mu_1+\mu_2+2)
\\
\phantom{}
\times
(p-x+\mu_1+\mu_2+1)(x-\mu_3)(2 p -x +\mu_1+\mu_2+\mu_3+2).
\end{gather*}
The eigenvalues of $A_1$ have the form:
\begin{gather}\label{eq:A1eig}
2 A_1
=\hbar^2 J (J + 1)  - \hbar^2 (\mu_1^2+\mu_2^2+\mu_3^2) + \frac{3}{4}\hbar^2,
\end{gather}
where $J=2 p +  \mu_1+ \mu_2+  \mu_3 + \frac{3}{2} $.

The eigenvalues of the operator $A_1$ are given by  the formula~(\ref{eq:A1eig}), where $p\ge m$.  The eigenvalues are calculated using the method described in Section~\ref{sec:TwoDim} for the subalgebra generated by the operators $A_2$, $B_2$, $A_1$.

Let now consider the subalgebra generated by the  operators $A_1$, $B_1$, $A_2$, $H$,  see equations~(\ref{eq:graph1}) and~(\ref{eq:coef1}).  The coef\/f\/icients~(\ref{eq:coef1}) of this algebra contain the eigenvalues of the operator~$A_2$, which was calculated previously{\samepage
\begin{gather*}
A_2=\frac{\hbar^2}{2}   (2 m + \mu_1+ \mu_2 + 1   )^2-\frac{\hbar^2}{2} \left( \mu_1^2+\mu_2^2\right)+\frac{\hbar^2}{4},
\end{gather*}
where $m=x$ or $m=p-x$ with $x=0,\ldots , p$ and $p\ge m$.}

The eigenvalues of the of the operator $A_1$, using the theory of Section~\ref{sec:TwoDim}  for the subalgebra generated by the operators $A_1$, $B_1$, $A_2$, $H$ is given by the formula
\begin{gather*}
A_1(y)=\frac{\gamma}{2} \left((y+v)^2 -\frac 14-\frac{\epsilon}{\gamma^2}\right).
\end{gather*}
From (\ref{eq:coef1}) we have that
\[
\gamma= 4 \hbar^2, \qquad \epsilon= 2\hbar^4 \left(2\mu_1^2+2\mu_2^2+2\mu_3^2-3\right).
\]
This formula should coincide with the formula calculated by equation (\ref{eq:A1eig}), therefore
\begin{gather*}
y=p\qquad  \mbox{and}  \qquad v=\frac{1}{2} (2+\mu_1+\mu_2+\mu_3).
\end{gather*}

The structure function $\Phi(v,y)$ has the following form
\begin{gather*}
\Phi(v,y)=3\cdot 2^{18} \hbar^{16} \left( 2\hbar^2 \mu^2 + (4(v+y)-3)^2 H \right)\left( 2\hbar^2 \mu^2 + (4(v+y)-1)^2 H \right)\\
\hphantom{\Phi(v,y)=}{}
\left( 8\hbar^2 (v+y)^2-8 \hbar^2 (v+y)(\mu_3+1)- 2 \hbar^2 \left(\mu_1^2+\mu_2^2-\mu_3^2-\frac{1}{2}\right)\right.\\
\left.\hphantom{\Phi(v,y)=}{}
+4\hbar^2 \left(\mu_3+\frac{1}{2}\right)-4 A_2 \right) \bigg( 8\hbar^2 (v+y)^2+8 \hbar^2 (v+y)(\mu_3-1)\\
\hphantom{\Phi(v,y)=}{}
- 2 \hbar^2 \left(\mu_1^2+\mu_2^2-\mu_3^2-\frac{1}{2}\right)-4\hbar^2 \left(\mu_3-\frac{1}{2}\right)-4 A_2 \bigg),
\end{gather*}
where $y=m,\ldots,q$.
The f\/inal form of the function $\Phi(v,y)$ with the above substitutions of $A_2$ and $v$ is
\begin{gather*}
\Phi(v,y)=3 \cdot 2^{24} \hbar^{20}(y-m)(1+m+y+\mu_1+\mu_2)(y-m+\mu_3)
(1+m+y+\mu_1+\mu_2+\mu_3)\\
\hphantom{\Phi(v,y)=}{}\times \left( 2\hbar^2\mu^2+ (1+4 y +2\mu_1+2\mu_2+2\mu_3)^2 H \right)\\
\hphantom{\Phi(v,y)=}{}\times
\left( 2\hbar^2\mu^2+ ( 3+4 y +2\mu_1+2\mu_2+2\mu_3)^2 H \right ).
\end{gather*}
 The condition for calculating the energy eigenvalues is
\[
\Phi(v,m)=0 , \qquad \Phi(v, q+1)=0,  \qquad \Phi(v,y)>0 \qquad \mbox{for} \quad y=m,  m+1,\ldots, q.
\]
The energy eigenvalues are calculated using the above relations
\begin{gather*}
H=-\frac{\hbar^2 \mu^2}{2 (\frac{5}{2}+2 q +\mu_1+\mu_2+\mu_3 )^2},\qquad H=-\frac{\hbar^2 \mu^2}{2 (\frac{5}{2}+2 q+1 +\mu_1+\mu_2+\mu_3 )^2},
\end{gather*}
where $ q=0,1,\ldots$.

\section{Discussion}
Using pure algebraic methods of~\cite{Das01}, we can calculate the energy eigenvalues of the nondegenerate three-dimensional Kepler--Coulomb system, which is discussed be Verrier and Evans~\cite{VerEvans08}.

This method can be applied to other three-dimensional nondegenerate superintegrable systems and it is the object of current investigation. The multidimensional ternary quadratic algebra is an algebra generated by the operators $S_i$ with $i=1, 2, \ldots, n$ satisfying the relations
\[
\left[ S_i, \left[ S_j,  S_k\right] \right]= \sum\limits_{r\le s}  d^{rs}_{ijk} \left\{ S_r, S_s\right\} +
 \sum\limits_{r} c^{r}_{ijk} S_r + f_{ijk}.
\]
The structure constants $ d^{rs}_{ijk}$, $c^{r}_{ijk}$, $f_{ijk}$ should obey to complicated restrictions, due the Jacobi kind relations for the quadratic algebra. The study of this kind of algebras, which describe many multidimensional superintegrable systems is an interesting mathematical topic, which is not yet been explored.

\appendix

\section{Appendix: Ternary quadratic algebra}\label{appendix}\vspace{-8mm}

\begin{gather}
[[A_1,B_1],A_2]=[[A_2,B_2],A_1]=[[A_1,F],B_2]= 0,\nonumber\\
[A_1,[A_1,B_1]]=-16 \hbar^2 H A_1^2 +4\hbar^2 \{A_1,B_1\}+ \left(16 \hbar ^2 A_2 H  -2 \hbar ^4  \left(4 \mu _1^2+4 \mu _2^2+12 \mu _3^2-5\right) H \right.\nonumber\\
 \left.
 \phantom{[A_1,[A_1,B_1]]=}{}
 -4 \hbar ^6 \mu ^2 \right) A_1 + \hbar ^4 \left(4 \mu _1^2+4 \mu _2^2+4 \mu _3^2-6\right) B_1 + 2 \hbar ^8 \mu ^2 \left(1-2 \mu _3^2\right) +4 \hbar ^6 \mu ^2 A_2\nonumber\\
 \phantom{[A_1,[A_1,B_1]]=}{}
 + \hbar ^6 \left(1-2 \mu _3^2 \left(4 \mu _1^2+4 \mu _2^2+4 \mu _3^2-5\right)\right) H\nonumber\\
 \phantom{[A_1,[A_1,B_1]]=}{}
  +2 \hbar ^4
   \left(4 \mu _1^2+4 \mu _2^2+4 \mu _3^2-5\right) A_2 H,\label{eq:alg1a}
\\
[B_1,[A_1,B_1]]=-4\hbar^2 B_1^2 +16\hbar^2 H \{A_1,B_1\} + 16\hbar^4 (5-4 \mu_3^2) H^2 A_1\nonumber\\
\phantom{[B_1,[A_1,B_1]]=}{}
 -\left(16 \hbar ^2 A_2 H  -2 \hbar ^4  \left(4 \mu _1^2+4 \mu _2^2+12 \mu _3^2-5\right) H  -4 \hbar ^6 \mu ^2 \right) B_1\nonumber\\
\phantom{[B_1,[A_1,B_1]]=}{}
 -2 \hbar^6  \left(16 \mu _3^4+8 \left(2 \mu _1^2+2 \mu _2^2-3\right) \mu _3^2-12 \mu _1^2-12 \mu _2^2+1\right) H^2\nonumber\\
\phantom{[B_1,[A_1,B_1]]=}{}
 +4  \hbar ^8 \mu ^2 \left(3-4 \mu _3^2\right) H-32 \hbar ^4 A_2 H^2,\label{eq:alg1b}
\\
[A_2,[A_2,B_2]]= 4\hbar^2 A_2^2 + 4\hbar^2 \{A_2,B_2\}+\left ( -4\hbar^2 A_1 +\hbar^4 \left( 4\mu_1^2-3\right)\right) A_2 \nonumber\\
\hphantom{[A_2,[A_2,B_2]]=}{} + \hbar^4\left(\hbar ^2 \mu _2^2-\frac{1}{4} \left(\hbar ^2+16 \mu _1^2-12\right)\right) A_1-\frac{\hbar ^6}{4}  \left(4 \mu _1^2-1\right)\nonumber\\
\hphantom{[A_2,[A_2,B_2]]=}{} + 2 \hbar ^4 \left(2 \mu _1^2+2 \mu _2^2-3\right) B_2,\label{eq:alg2a}\\
[B_2,[A_2,B_2]]= -4\hbar^2 B_2^2 -4\hbar^2\{A_2,B_2\}-2 \hbar ^4 \left(2 \mu _1^2+2 \mu _3^2-3\right) A_2\nonumber\\
\phantom{[B_2,[A_2,B_2]]=}{}
-\left ( -4\hbar^2 A_1 +\hbar^4 \left( 4\mu_1^2-3\right)\right) B_2 + \frac{1}{4} \hbar ^6\left(4 \mu _1^2-1\right) \nonumber\\
\phantom{[B_2,[A_2,B_2]]=}{}
+\frac{1}{4} \hbar ^4 \left(-4 \mu _3^2 \hbar ^2+\hbar ^2+16 \mu _1^2-12\right) A_1,\label{eq:alg2b}
\\
[A_1,[A_1,F]]= \left( 16\hbar^2 B_2 H-2\hbar^4 \left(4\mu_1^2+12\mu_2^2+4\mu_3^2-5\right) H -4 \hbar^6 \mu^2 \right) A_1\nonumber\\
\hphantom{[A_1,[A_1,F]]=}{}
-16 \hbar^2 H A_1^2 + 4\hbar^2 \{A_1,F\}+ \hbar^4\left( 4 \mu_1^2+4 \mu_2^2+4 \mu_3^2 -6  \right) F  +4 \hbar^6 \mu ^2 B_2\nonumber\\
\hphantom{[A_1,[A_1,F]]=}{}
 + \hbar ^6 \left(1-2 \mu _2^2 \left(4 \mu _1^2+4 \mu _2^2+4 \mu _3^2-5\right)\right) H +2
    \hbar ^4 \left(4 \mu _1^2+4 \mu _2^2+4 \mu _3^2-5\right) B_2 H\nonumber\\
\hphantom{[A_1,[A_1,F]]=}{}
    + 2 \hbar ^8 \mu ^2 \left(1-2 \mu _2^2\right),\label{eq:alg3a}\\
[F,[A_1,F]]=-2  \hbar ^6 \left(4 \left(4 \mu _2^2-3\right) \mu _1^2-12 \mu _3^2+8 \mu _2^2 \left(2 \mu _2^2+2 \mu
   _3^2-3\right)+1\right) H^2\nonumber\\
\hphantom{[F,[A_1,F]]=}{}
   -\left( 16\hbar^2 B_2 H-2\hbar^4 \left(4\mu_1^2+12\mu_2^2+4\mu_3^2-5\right) H -4 \hbar^6 \mu^2 \right) F-4 \hbar^2 F^2 \label{eq:alg3b} \\
\hphantom{[F,[A_1,F]]=}{}
   +16 \hbar^2 H \{A_1,F\}+ 16 \hbar^4 \left( 5-4\mu_2^2\right) H^2 A_1 + 4 \hbar ^8  \mu ^2 \left(3-4 \mu _2^2\right) H  -32 \hbar^4 B_2 H^2, \nonumber\\
[A_1,[B_1,B_2]]=[[A_1,B_1],B_2]= \hbar ^6 \left(8 \mu _3^4+\left(8 \mu _1^2-6\right) \mu _3^2+\mu _2^2 \left(8 \mu _3^2-4\right)-1\right) H
\nonumber\\
\phantom{[A_1,[B_1,B_2]]=}{}
+ 2 \mu ^2 \hbar ^8 \left(2 \mu _3^2-1\right)+16 \hbar^2 A_1^2 H-16\hbar^2 A_1 A_2 H\nonumber\\
\phantom{[A_1,[B_1,B_2]]=}{}
+2 \hbar ^4 \left(4 \mu _1^2+4 \mu _2^2+12 \mu _3^2-5\right) A_1 H-2 \hbar ^4 \left(4 \mu _1^2+4 \mu _2^2+4 \mu _3^2-3\right) A_2 H
\nonumber\\
\phantom{[A_1,[B_1,B_2]]=}{}
+4 \hbar^4 B_2 H-2\hbar^2 \{A_1,B_1\}-4 \hbar ^6 \mu ^2  A_2-4 \hbar ^4 \left(\mu _1^2+\mu _3^2-1\right) B_1\nonumber\\
\phantom{[A_1,[B_1,B_2]]=}{}
+2 \hbar ^4 \left(1-2 \mu _3^2\right) F-2\hbar^2 \{B_1,B_2\}-2\hbar^2 \{A_1,F\}+2 \hbar^2 \{A_2,F\}+4 \mu ^2 \hbar ^6 A_1,\nonumber\\
[[B_1,B_2],A_2]=[[A_2,B_2],B_1]=-4 \hbar ^6 \mu ^2  A_1+4 \hbar ^6 \mu ^2  A_2 +4 \hbar ^6 \mu ^2 B_2 +\hbar^4 B_1 +2 \hbar^4 F
\nonumber\\
\hphantom{[[B_1,B_2],A_2]=}{}
-16 \hbar^2 A_1^2 H+16 \hbar^2 A_1 A_2 H+16 \hbar^2 A_1 B_2 H
\nonumber\\
\hphantom{[[B_1,B_2],A_2]=}{}
-2 \hbar ^4 \left(4 \mu _1^2+4 \mu _2^2+4 \mu _3^2+1\right) A_1 H
-2\hbar^2 \{B_1,B_2\} +2 \hbar^2\{ A_1,F\} \nonumber\\
\hphantom{[[B_1,B_2],A_2]=}{}
-2\hbar^2 \{A_2,F\} -\frac{1}{2} \hbar ^6 \left(4 \mu _1^2+4 \mu _2^2+4 \mu _3^2+1\right) H-4\hbar^2 A_2 B_1  -\hbar ^8 \mu ^2\nonumber\\
\hphantom{[[B_1,B_2],A_2]=}{}
 +2 \hbar ^4 \left(4 \mu _1^2+4 \mu _2^2+4 \mu _3^2-1\right) A_2 H+2 \hbar ^4 \left(4 \mu _1^2+4 \mu _2^2+4 \mu _3^2-3\right) B_2 H\nonumber\\
 \hphantom{[[B_1,B_2],A_2]=}{}
 +2 \hbar^2 \{A_1,B_1\},
\nonumber\\
[[A_1,F],A_2]=[[A_2,F],A_1]= \hbar ^6 \left(8 \mu _2^4+\left(8 \mu _1^2+8 \mu _3^2-6\right) \mu _2^2-4 \mu _3^2-1\right) H\nonumber\\
\hphantom{[[A_1,F],A_2]=}{}
+ 2 \hbar ^8 \mu ^2  \left(2 \mu _2^2-1\right)+16 \hbar^2 A_1^2 H-16 \hbar^2 A_1 B_2 H\nonumber\\
\hphantom{[[A_1,F],A_2]=}{}
+2 \hbar ^4 \left(4 \mu _1^2+12 \mu _2^2+4 \mu _3^2-5\right) A_1 H+4 \hbar^4 A_2 H\nonumber\\
\hphantom{[[A_1,F],A_2]=}{}
-2 \hbar ^4 \left(4 \mu _1^2+4 \mu _2^2+4 \mu _3^2-3\right) B_2 H-2\hbar^2 \{A_1,B_1\}+2\hbar^2 \{B_1,B_2\}\nonumber\\
\hphantom{[[A_1,F],A_2]=}{}
-2\hbar^2 \{A_1,F\}+4 \hbar^6 \mu^2 A_1+2 \hbar ^4 \left(1-2 \mu _2^2\right) B_1-4 \hbar ^6 \mu ^2  B_2\nonumber\\
\hphantom{[[A_1,F],A_2]=}{}
-4 \hbar ^4 \left(\mu _1^2+\mu _2^2-1\right) F-2 \hbar^2 \{A_2,F\},
\nonumber\\
[[A_2,F],B_1]=[[B_1,F],A_2]= 64 \hbar^2 A_1^2 H^2-64 \hbar^2 A_1 A_2 H^2 -64 \hbar^2 A_1 B_2 H^2 -8 \hbar^2 \{A_1,B_1\} H \nonumber\\
\phantom{[[A_2,F],B_1]=}{}
+ 16 \hbar^2 A_2 B_1 H + 8 \hbar^2 \{B_1, B_2\} H -8 \hbar^2 \{A_1,F\} H +8 \hbar^2 \{A_2,F\} H \nonumber\\
\phantom{[[A_2,F],B_1]=}{}
+ 8 \hbar ^4 \left(4 \mu _1^2+4 \mu _2^2+4 \mu _3^2+1\right) A_1 H^2-8 \hbar ^4 \left(4 \mu _1^2+4 \mu _2^2+4 \mu _3^2-1\right) A_2 H^2
 \nonumber\\
\phantom{[[A_2,F],B_1]=}{}
 -8 \hbar ^4 \left(4 \mu _1^2+4 \mu _2^2+4 \mu _3^2-3\right) B_2 H^2 + 16 \hbar ^6 \mu ^2 A_1 H -16 \hbar ^6 \mu ^2 A_2 H
 \nonumber\\
\phantom{[[A_2,F],B_1]=}{}
 -16 \hbar ^6 \mu ^2 B_2 H -8 \hbar^4 F H -4 \hbar^4 B_1 H + 2 \hbar ^6 \left(4 \mu _1^2+4 \mu _2^2+4 \mu _3^2+1\right) H^2
 \nonumber\\
\phantom{[[A_2,F],B_1]=}{}
 + 4 \hbar^8 \mu^2 H,\nonumber\\
[[A_2,B_2],F]=[[A_2,F],B_2]= 16 \hbar^2 A_1^2 H -16 \hbar^2 A_1 A_2 H -16 \hbar^2 A_1 B_2 H
\nonumber\\
\phantom{[[A_2,B_2],F]=}{}
+2 \hbar ^4 \left(4 \mu _1^2+4 \mu _2^2+4 \mu _3^2+1\right) A_1 H -2 \hbar ^4 \left(4 \mu _1^2+4 \mu _2^2+4 \mu _3^2-3\right) A_2 H
 \nonumber\\
\phantom{[[A_2,B_2],F]=}{}
 -2 \hbar ^4 \left(4 \mu _1^2+4 \mu _2^2+4 \mu _3^2-1\right) B_2 H -2 \hbar^2 \{A_1,B_1\} +2 \hbar^2 \{B_1,B_2\}
 \nonumber\\
\phantom{[[A_2,B_2],F]=}{}
  -2 \hbar^2 \{A_1, F\} +2 \hbar^2 \{A_2,F\} +4 \hbar^2 F B_2 + 4 \hbar ^6 \mu ^2  A_1 -4 \hbar ^6\mu ^2  A_2 -4 \hbar ^6\mu ^2  B_2
 \nonumber\\
\phantom{[[A_2,B_2],F]=}{}
   -2 \hbar^4 B_1 -\hbar^4 F + \frac{1}{2} \hbar ^6 \left(4 \mu _1^2+4 \mu _2^2+4 \mu _3^2+1\right) H + \hbar^8\mu^2,
\nonumber\\
[[B_1,B_2],F]=[[B_1,F],B_2]= -64 \hbar^2 A_1^2 H^2 +64 \hbar^2 A_1 A_2 H^2 +64 \hbar^2 A_1 B_2 H^2
\nonumber\\
\hphantom{[[B_1,B_2],F]=}{}
+8 \hbar^2 \{A_1,B_1\} H +8 \hbar^2\{ A_1,F\} H -8 \hbar^2 \{A_2,F\} H -8 \hbar^2 \{B_1, B_2\} H
\nonumber\\
\hphantom{[[B_1,B_2],F]=}{}
-16 \hbar^2 B_2 F H-8 \hbar ^4 \left(4 \mu _1^2+4 \mu _2^2+4 \mu _3^2+1\right) A_1 H^2
\nonumber\\
\hphantom{[[B_1,B_2],F]=}{}
 + 8 \hbar ^4 \left(4 \mu _1^2+4 \mu _2^2+4 \mu _3^2-3\right) A_2 H^2 + 8 \hbar ^4 \left(4 \mu _1^2+4 \mu _2^2+4 \mu _3^2-1\right) B_2 H^2
\nonumber\\
\hphantom{[[B_1,B_2],F]=}{}
 -16 \hbar ^6 \mu ^2 A_1 H + 16 \hbar ^6 \mu ^2  A_2 H + 16 \hbar ^6 \mu ^2  B_2 H +4 \hbar^4 F H + 8 \hbar^4 B_1 H
\nonumber\\
\hphantom{[[B_1,B_2],F]=}{}
  -2 \hbar ^6 \left(4 \mu _1^2+4 \mu _2^2+4 \mu _3^2+1\right) H^2 -4 \hbar ^8 \mu ^2 H,
\nonumber\\
[[B_1,F],A_1]=8 \hbar^2\{A_1, B_1\} H-8 \hbar^2\{ B_1,B_2\} H+8 \hbar^2 \{A_2,F\} H-8 \hbar^2\{A_1,F\} H
\nonumber\\
\phantom{[[B_1,F],A_1]=}{}
-16 \hbar^4 A_2 H^2 +16 \hbar^4 B_2 H^2 + 8 \hbar ^4 \left(1-2 \mu _3^2\right) F H + 8 \hbar ^4 \left(2 \mu _2^2-1\right) B_1 H
\nonumber\\
\phantom{[[B_1,F],A_1]=}{}
+ 16 \hbar ^6 \left(\mu _3^2-\mu _2^2\right) H^2,
\nonumber\\
[[A_1,B_1],F]= 64 \hbar^2 A_1^2 H^2 -64 \hbar^2 A_1 A_2 H^2 -64 \hbar^2 A_1 B_2 H^2+8 \hbar^2 \{B_1,B_2\} H
 \nonumber\\
 \phantom{[[A_1,B_1],F]=}{}
-8 \hbar^2 \{A_1,B_1\} H + 8 \hbar ^4 \left(4 \mu _1^2+4 \mu _2^2+4 \mu _3^2+5\right) A_1 H^2
\nonumber\\
 \phantom{[[A_1,B_1],F]=}{}
-8 \hbar ^4 \left(4 \mu _1^2+4 \mu _2^2+4 \mu _3^2+1\right) A_2 H^2 -8 \hbar ^4 \left(4 \mu _1^2+4 \mu _2^2+4 \mu _3^2+3\right) B_2 H^2
\nonumber\\
 \phantom{[[A_1,B_1],F]=}{}
+ 4 \hbar ^4 \left(1-4 \mu _2^2\right) B_1 H +16 \hbar ^6  \mu ^2 A_1 H -16 \hbar ^6 \mu ^2  A_2 H -16 \hbar ^6 \mu ^2 B_2 H
\nonumber\\
 \phantom{[[A_1,B_1],F]=}{}
-4 \hbar ^4 F H +2 \hbar^2 \{B_1,F\} + 2 \hbar ^6 \left(4 \mu _1^2+12 \mu _2^2+4 \mu _3^2-1\right) H^2 + 4 \hbar^8 \mu^2 H,
\nonumber\\
[[A_1,F],B_1]= 64 \hbar^2 A_1^2 H^2-64 \hbar^2 A_1 A_2 H^2 -64 \hbar^2 A_1 B_2 H^2
\nonumber\\
\phantom{[[A_1,F],B_1]=}{}
+ 8 \hbar ^4 \left(4 \mu _1^2+4 \mu _2^2+4 \mu _3^2+5\right) A_1 H^2 -8 \hbar ^4 \left(4 \mu _1^2+4 \mu _2^2+4 \mu _3^2+3\right) A_2 H^2
\nonumber\\
\phantom{[[A_1,F],B_1]=}{}
-8 \hbar ^4 \left(4 \mu _1^2+4 \mu _2^2+4 \mu _3^2+1\right) B_2 H^2 -8 \hbar^2 \{A_1, F\} H +8\hbar^2 \{ A_2,F\} H
\nonumber\\
\phantom{[[A_1,F],B_1]=}{}
+16 \hbar ^6 \mu ^2 A_1 H-16 \hbar ^6 \mu ^2 A_2 H-16 \hbar ^6 \mu ^2 B_2 H + 4 \hbar ^4 \left(1-4 \mu _3^2\right) F H
\nonumber\\
\phantom{[[A_1,F],B_1]=}{}
-4 \hbar ^4 B_1 H +2 \hbar^2 \{B_1, F\} + 2 \hbar ^6 \left(4 \mu _1^2+4 \mu _2^2+12 \mu _3^2-1\right) H^2 +4 \hbar^8 \mu^2 H,
\nonumber\\
[[B_1,F],F]= 64 \hbar^2 B_2 F H^2 -32\hbar^2 \{A_1,F\} H^2 + 32 \hbar^2 \{ A_2,F\} H^2 -128 \hbar^4 A_1 H^3
\nonumber\\
\phantom{[[B_1,F],F]=}{}
+128 \hbar^4 A_2 H^3 +128 \hbar^4 B_2 H^3 -16 \hbar ^4 \left(3-4 \mu _2^2\right)  B_1 H^2 -8 \hbar^2 \{B_1, F\} H
\nonumber\\
\phantom{[[B_1,F],F]=}{}
 + 16 \hbar ^6 \left(1-4 \mu _2^2\right) H^3,
\nonumber\\
[[B_1,F],B_1]= 32 \hbar^2\{A_1,B_1\} H^2 -64 \hbar^2 B_1 A_2 H^2 -32 \hbar^2 \{B_1, B_2\} H^2 +128 \hbar^4 A_1 H^3
\nonumber\\
\phantom{[[B_1,F],B_1]=}{}
-128 \hbar^4 A_2 H^3 -128 \hbar^4 B_2 H^3 + 16 \hbar ^4 \left(3-4 \mu _3^2\right) F H^2 + 8 \hbar^2 \{B_1, F\} H
\nonumber\\
\phantom{[[B_1,F],B_1]=}{}
- 16 \hbar ^6 \left(1-4 \mu _3^2\right) H^3,
\nonumber\\
[[A_2,F],F]= -8 \hbar^2 \{A_1,F\} H -8 \hbar^2 \{A_2,F\} H + 16 \hbar ^4 \left(4 \mu _2^2-3\right) A_1 H^2 -32 \hbar^4 A_2 H^2
\nonumber\\
\phantom{[[A_2,F],F]=}{}
+ 2 \hbar ^6 \left(16 \mu _2^4+16 \left(\mu _3^2-1\right) \mu _2^2-12 \mu _3^2+4 \mu _1^2 \left(4 \mu _2^2-3\right)-1\right) H^2+2 \hbar^2 \{B_1,F\}
\nonumber\\
\phantom{[[A_2,F],F]=}{}
 +4 \hbar^2 F^2 -2 \hbar ^4 \left(4 \mu _1^2+12 \mu _2^2+4 \mu _3^2-5\right) F H +4 \hbar ^4 \left(3-4 \mu _2^2\right) B_1 H -4 \hbar ^6 \mu ^2 F
\nonumber\\
\phantom{[[A_2,F],F]=}{}
  + 4 \hbar ^8 \mu ^2  \left(4 \mu _2^2-3\right) H,
\nonumber\\
[[F,A_2],A_2]= -\frac{1}{2} \hbar ^6 \left(16 \mu _2^4+16 \left(\mu _3^2-1\right) \mu _2^2+4 \mu _1^2 \left(4 \mu _2^2-1\right)-3 \left(4 \mu _3^2+1\right)\right) H
\nonumber\\
\phantom{[[F,A_2],A_2]=}{}
-16 \hbar^2 A_1 A_2 H +4 \hbar^2 A_2 B_1 + 4\hbar^2 \{A_2,F\}+4 \hbar ^4 \left(3-4 \mu _2^2\right) A_1 H
\nonumber\\
\phantom{[[F,A_2],A_2]=}{}
-2 \hbar ^4 \left(4 \mu _1^2+4 \mu _2^2+4 \mu _3^2+1\right) A_2 H + 2 \hbar ^4 \left(2 \mu _1^2+2 \mu _2^2-3\right) F
\nonumber\\
\phantom{[[F,A_2],A_2]=}{}
+ \hbar ^4 \left(4 \mu _2^2-3\right) B_1 -4 \hbar ^6 \mu ^2 A_2+ \hbar ^8 \mu ^2  \left(3-4 \mu _2^2\right),
\nonumber\\
[[B_2,B_1],B_1]=2 \hbar ^6 \left(16 \mu _3^4-16 \mu _3^2+4 \mu _1^2 \left(4 \mu _3^2-3\right)+4 \mu _2^2 \left(4 \mu _3^2-3\right)-1\right) H^2 -32 \hbar^4 B_2 H^2
\nonumber\\
\phantom{[[B_2,B_1],B_1]=}{}
 +2 \hbar^2\{ B_1, F\} -2 \hbar ^4 \left(4 \mu _1^2+4 \mu _2^2+12 \mu _3^2-5\right) B_1 H +4 \hbar^2 B_1^2
\nonumber\\
\phantom{[[B_2,B_1],B_1]=}{}
 + 4 \hbar ^4 \left(3-4 \mu _3^2\right) F H -4 \hbar ^6 \mu ^2  B_1+ 4 \mu ^2 \hbar ^8 \left(4 \mu _3^2-3\right) H-8 \hbar^2 \{A_1,B_1\} H
\nonumber\\
\phantom{[[B_2,B_1],B_1]=}{}
  -8 \hbar^2 \{B_1, B_2\} H + 16 \hbar ^4 \left(4 \mu _3^2-3\right) A_1 H^2,
\nonumber\\
[[B_2,B_1],B_2]= \frac{1}{2} \hbar ^6 \left(16 \mu _3^4-16 \mu _3^2+4 \mu _2^2 \left(4 \mu _3^2-3\right)+4 \mu _1^2 \left(4 \mu _3^2-1\right)-3\right) H +16 \hbar^2 A_1 B_2 H
\nonumber\\
\phantom{[[B_2,B_1],B_2]=}{}
- 4\hbar^2 \{B_1,B_2\}+4 \hbar ^4 \left(4 \mu _3^2-3\right) A_1 H+2 \hbar ^4 \left(4 \mu _1^2+4 \mu _2^2+4 \mu _3^2+1\right) B_2 H
\nonumber\\
\phantom{[[B_2,B_1],B_2]=}{}
-2 \hbar ^4 \left(2 \mu _1^2+2 \mu _3^2-3\right) B_1-4 \hbar^2 B_2 F +\hbar ^4 \left(3-4 \mu _3^2\right) F + 4 \hbar ^6 \mu ^2 B_2
\nonumber\\
\phantom{[[B_2,B_1],B_2]=}{}
+  \mu ^2 \hbar ^8 \left(4 \mu _3^2-3\right),
\nonumber\\
[A_1,[A_1,F]]= \left( 16\hbar^2 B_2 H-2\hbar^4 \left(4\mu_1^2+12\mu_2^2+4\mu_3^2-5\right) H -4 \hbar^6 \mu^2 \right) A_1-16 \hbar^2 H A_1^2
\nonumber\\
\phantom{[A_1,[A_1,F]]=}{}
 + 4\hbar^2 \{A_1,F\}+ \hbar^4\left( 4 \mu_1^2+4 \mu_2^2+4 \mu_3^2 -6  \right) F  +4 \hbar^6 \mu ^2 B_2
\nonumber\\
\phantom{[A_1,[A_1,F]]=}{}
 + \hbar ^6 \left(1-2 \mu _2^2 \left(4 \mu _1^2+4 \mu _2^2+4 \mu _3^2-5\right)\right) H +2
    \hbar ^4 \left(4 \mu _1^2+4 \mu _2^2+4 \mu _3^2-5\right) B_2 H
\nonumber\\
\phantom{[A_1,[A_1,F]]=}{}
    + 2 \hbar ^8 \mu ^2 \left(1-2 \mu _2^2\right),
\nonumber\\
[F,[A_1,F]]=-2  \hbar ^6 \left(4 \left(4 \mu _2^2-3\right) \mu _1^2-12 \mu _3^2+8 \mu _2^2 \left(2 \mu _2^2+2 \mu
   _3^2-3\right)+1\right) H^2
\nonumber\\
\phantom{[F,[A_1,F]]=}{}
   -\left( 16\hbar^2 B_2 H-2\hbar^4 \left(4\mu_1^2+12\mu_2^2+4\mu_3^2-5\right) H -4 \hbar^6 \mu^2 \right) F-4 \hbar^2 F^2
\nonumber\\
\phantom{[F,[A_1,F]]=}{}
    +16 \hbar^2 H \{A_1,F\}+ 16 \hbar^4 \left( 5-4\mu_2^2\right) H^2 A_1 + 4 \hbar ^8  \mu ^2 \left(3-4 \mu _2^2\right) H  -32 \hbar^4 B_2 H^2.\nonumber
\end{gather}

\pdfbookmark[1]{References}{ref}
\LastPageEnding

\end{document}